# Optical patterning of magnetic domains and defects in ferromagnetic liquid crystal colloids

Andrew J. Hess,[1] Qingkun Liu,[1] and Ivan I. Smalyukh[1,2,3]*

[1]Department of Physics, University of Colorado, Boulder, CO 80309, USA
[2]Materials Science and Engineering Program, Department of Electrical, Computer and Energy Engineering, Liquid Crystal Materials Research Center, University of Colorado, Boulder, CO 80309, USA
[3]Renewable and Sustainable Energy Institute, National Renewable Energy Laboratory and University of Colorado, Boulder, CO 80309, USA

*Corresponding author: ivan.smalyukh@colorado.edu

*A promising approach in designing composite materials with unusual physical behavior combines solid nanostructures and orientationally ordered soft matter at the mesoscale. Such composites not only inherit properties of their constituents but also can exhibit emergent behavior, such as ferromagnetic ordering of colloidal metal nanoparticles forming mesoscopic magnetization domains when dispersed in a nematic liquid crystal. Here we demonstrate the optical patterning of domain structures and topological defects in such ferromagnetic liquid crystal colloids, which allows for altering their response to magnetic fields. Our findings reveal the nature of the defects in this soft matter system which is different as compared to non-polar nematics and ferromagnets alike.*

Liquid crystal (LC) colloids attract a considerable amount of scientific interest driven by the richness of their fundamental physics behavior and potential for applications [1-4]. This is especially the case for dispersions of anisotropic metal nanoparticles, such as rods and platelets, in a nematic host, in which particle-medium interactions result in spontaneous orientational ordering of the inclusions [3-7]. This ordering can be pre-engineered to exhibit both positive and negative values of the scalar order parameter describing self-organization of dispersed nanoparticles [3,6] and can be both non-polar and polar in nature [4,7]. Ferromagnetic liquid crystal colloids (FLCCs) with polar ordering of magnetic dipoles of individually dispersed nanoparticles, similar to many other ferromagnetic systems [8], exhibit a magnetic domain structure with domains having different orientations of magnetization **M** [4]. The response of individual magnetic domains to external fields is dependent on the relative orientation of **M** within the domains [7] and the applied magnetic field **B**. Although control of the magnetic domain spatial distribution is important from both fundamental and applications-related standpoints [8], achieving it is typically difficult. Here we show that this domain structure in FLCCs can be robustly controlled using low-intensity unstructured light.

In this letter, we demonstrate light-directed spatial patterning of magnetization monodomains with a well-defined direction of $\mathbf{M}$ that can be guided to orient along one of the two anti-parallel directions of uniform non-polar ordering of rod-like LC molecules described by the far-field director $\mathbf{n}_0$. This is achieved through first driving a light-induced order-disorder melting transition of the nematic host within the illuminated parts of the sample and then quenching it back to the ordered state in the presence of the magnetic field. We also show that continuously varying patterns of the director field $\mathbf{n(r)}$ and magnetization field $\mathbf{M(r)}$ $\parallel\mathbf{n(r)}$ can be obtained via optical patterning of the surface boundary conditions. Both bulk-assisted alignment of $\mathbf{M(r)}$ and surface-mediated patterning of $\mathbf{n(r)}$ in this system are achieved using azobenzene-containing molecules [9,10-14] dispersed in the sample bulk or localized at the sample surfaces, respectively. Our studies reveal the nature of topological defects in this soft matter system with polar ordering and also show the means of engineering its response to weak external stimuli including magnetic fields and light.

Barium hexaferrite $BaFe_{11.5}Cr_{0.5}O_{19}$ ferromagnetic nanoplatelets were synthesized by the thermohydrodrate method [15]. 0.01 M of $Ba(NO_3)_2$, 0.045 M of $Fe(NO_3)_3$, and 0.005 M of $Cr(NO_3)_3$ were dissolved in deionized water and co-precipitated by 2.72 M of NaOH aqueous solution (all ingredients from Alfa Aesar) in a 25 mL teflon-lined autoclave. The solution was hydrothermally heated to 220 °C at 3 °C/min, held for 1 h, and then cooled down to room temperature. Precipitated powders were washed with 10 wt.% nitric acid and acetone and re-dispersed in 1 mL of water, yielding magnetically monodomain platelets of $\approx$10 nm in thickness and, on average, 195 nm in diameter and with magnetic moments $\approx$2.2 × $10^{-17}$ Am$^2$ orthogonal to their large-area faces at the maximum magnetic field of 800kA/m. To improve colloidal stability and define homeotropic boundary conditions for the LC molecular alignment and the director field $\mathbf{n(r)}$ at the platelets' surfaces [16], 100μL of 3 wt.% of this colloidal dispersion was added to 10mL of water adjusting the pH to be within 3-4 by nitric acid. Then, 1 mL of ethanol solution with 10 mg of 5 kDa silane-terminated polyethylene glycol (JemKem Technology) was added to the particle dispersion drop-wise so that the silane groups hydrolyzed and linked to OH groups at platelet surfaces. The dispersion was kept for 12h, followed by centrifugation at 14,600 rpm for 30 min, and then washed by ethanol twice.

To prepare photo-isomerizing FLCCs, which can transition from nematic to isotropic phase upon illumination, 95 μL of pentylcyanobiphenyl (5CB, ChengzhiYonghua Display Materials Co. Ltd.) was mixed with 5 μL of azobenzene-containing Beam 1205 LC (Beam Co.). Non-isomerizing FLCCs utilized for surface-enabled optical defect patterning were based on the pure 5CB host medium. Silane-PEG capped magnetic nanoplatelets dispersed in ethanol were then added to the LC while following dispersion procedures reported previously [5]. 5 μL of ethanol was added to 15 μL of the LC mixture to bring it to the isotropic phase, followed by adding 15 μL of 0.5-1 wt.% magnetic platelets dispersed in ethanol. The sample was kept at 90 °C for 3 h to fully evaporate the ethanol, yielding an excellent dispersion in the isotropic phase at no fields, and then was rapidly cooled to the nematic phase of the mixture while vigorously stirring. The ensuing FLCC was centrifuged at 2,000 rpm for 5 min to remove residual aggregations so that the final composite contained only well-dispersed platelets. The final fraction of magnetic platelets in the LC was varied within 0.05-0.1wt.%, as determined based on absorbance and magnetization values. Nanoplatelets exhibited spontaneous alignment with large-area faces orthogonal to $\mathbf{n(r)}$ and magnetic moments along $\mathbf{n(r)}$, as confirmed by measuring polarization-dependent absorbance [16] and probing response of their dilute dispersions to magnetic fields. Dispersions of ferromagnetic platelets in this LC mixture were stable at all used fields (up to 20 mT) exhibiting a facile response already at fields below 1 mT. Homeotropic glass cells with polydomain FLCCs and $\mathbf{M}$ pointing along one of the two anti-parallel directions along the vertical far-field director $\mathbf{n}_0$ were prepared using 1- or 0.17-mm thick glass plates treated with an aqueous solution of 0.1 wt. % N,N-dimethyl-N-octadecyl-3-aminopropyl-trimethoxysilyl chloride (DMOAP, AcrosOrganics) via dip-coating. The cell gap thickness of 30 µm or 60 µm was set by inter-spacing the glass plates at their edges with UV-curable optical adhesive (NOA-65, Norland Products) or epoxy containing silica spacers of corresponding diameters. Cell filling was done at room temperature using capillary action.  FLCC cells exhibit spontaneous formation of random magnetic domains of lateral dimensions typically comparable or somewhat larger than the cell thickness and dependent on the initialization field of 10-35 mT [4,16] (Fig. 1) as well as polar switching at low (~1mT) magnetic fields. To probe and control them, we used an Olympus polarizing optical microscope (POM) BX51, which was equipped with a heating stage (Instec STC200) mounted on a rotating microscope stage. An air-core solenoid was used to apply magnetic fields up to 20mT in the upward and downward vertical directions orthogonal to the cell substrates. Permanent magnets and metal-core solenoids integrated with a home-built POM were utilized to apply uniform fields up to 35 mT along different directions in the plane of a FLCC cell [17].

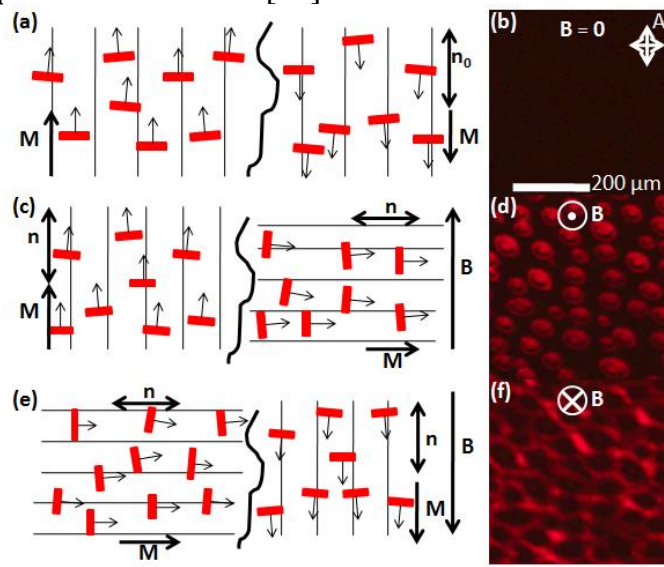

Fig. 1.Magnetic switching of polydomain FLCCs. (a) A schematic of spontaneously formed domains with $\mathbf{M}$ in one of the two anti-parallel directions along the non-

polar far-field director $n_0 \equiv -n_0$. Black lines depict the director $n$ while black arrows on the orange platelets represent their magnetic dipole moments. White double arrows marked with "P" and "A" show the crossed polarizer and analyzer, respectively. (b) A corresponding POM image with the domains invisible at no fields because of the homeotropic $n_0$. (c) A schematic of the same domains as shown in (a) but, after applying a uniform field ≈3 mT, $M$ rotates in domains where it initially was antiparallel to the applied field $B$ but stays intact in regions with initial $M \| B$. (d) A corresponding POM image. (e,f) When vertical $B$ is reversed as compared to (c,d), the complementary set of magnetic domains exhibit response, as shown schematically in (e) and seen by POM in (f). Red-light illumination in (b,d,f) helped to avoid unintended switching of photo-isomerizing FLCCs during POM imaging. The cell thickness is 30 μm.

To optically pattern ferromagnetic domains within initially polydomain FLCCs prepared as described above, the cell temperature was raised to ≈36 °C, right below the nematic to isotropic transition temperature ≈39 °C of the used mixture of 5CB and Beam 1205. This elevated temperature enables spatially controlled photo-isomerization leading to laterally localized phase transition in the illuminated areas (Fig. 2a). A magnetic field of ≈10 mT was applied using a solenoid placed directly over the cell such that its symmetry axis was coaxial with the optical axis of the microscope. The spatial pattern of the photo-isomerization-enabled nematic-isotropic transition (Fig. 2a) was defined using an LC microdisplay-based optical patterning system coupled to POM and described in detail elsewhere [11]. Broadband blue light was used to photo-induce the arbitrary patterns of localized isotropic phase within the cell containing a nematic dispersion of magnetic platelets. A field of ≈3 mT was then applied perpendicular to the sample plane as the photo-induced isotropic-phase region transitioned back to the ordered phase upon reduction of the temperature to 31 °C. This defined the same direction of $M$ within the entire sample area of illumination (Fig. 2b). The patterned monodomain, similar to spontaneously occurring domains with opposite directions of $M$, remains invisible between crossed polarizers of POM at no fields because of the homeotropic $n_0$ both within and around it. This photo-patterned ferromagnetic domain remains intact at fields $B \| M$ within it but responds uniformly to $B$ above ≈1 mT orientated anti-parallel to $M$, as evidenced by the change of brightness in POM (Fig. 2c-f), while the surrounding region around the domain can remain in a polydomain state or can be patterned to have the same opposite orientation of $M$. Control of parameters, such as the strength of the field applied while in the photo-induced isotropic state, allows for different degrees of patterning of $M$ and the relative fraction of patterned and spontaneous domains, which can be then assessed by applying ≈3 mT to elicit complementary domains within and around the photo-patterned region (Fig. 2c-h). In addition to homeotropic cells, we also achieved magnetic domains in glass cells with in-plane $M \| n_0$ by quenching them from photo-isomerization-enabled isotropic regions in the presence of in-plane field $B \| n_0$ selecting one of the two desired directions of $M \| B \| n_0$.

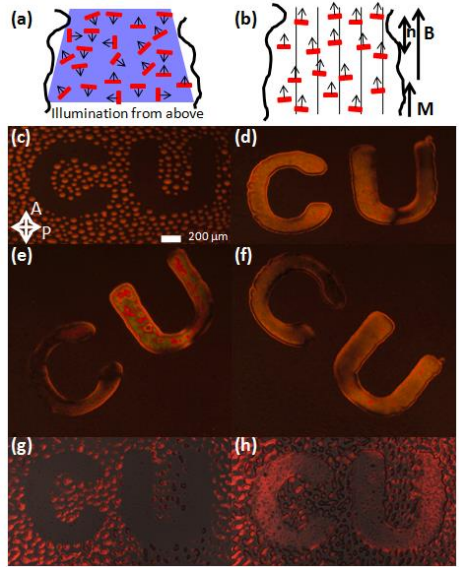

Fig. 2. Formation and switching of photo-patterned FLCC monodomains. (a) A cross-section schematic of a region of the cell illuminated with broadband blue light, in which the cis-trans isomerization of the azobenzene in Beam 1205 molecules within the LC host mixture causes an order-disorder transition in the illuminated areas to photo-induce a laterally localized isotropic-phase region. (b) When the patterning region is still in the photo-induced isotropic state, a uniform vertical $\mathbf{B}$ is applied while the cell is quenched back to the ordered state with $\mathbf{M}\|\mathbf{B}\|\mathbf{n_0}$. The surface anchoring-mediated mechanical coupling of nanoplatelet orientations to $\mathbf{n_0}$ then maintains this $\mathbf{M}$ within an arbitrarily shaped patterned monodomain. (c) POM of a magnetic domain shaped as "CU" defined in a cell with surrounding random naturally occurring domains of much smaller size visible at $\mathbf{B}\|\mathbf{M}$ of the "CU" domain. (d-f) The "CU" domain in a uniform surrounding observed in POM at different orientations with respect to crossed polarizers at $\mathbf{B}$ anti-parallel to $\mathbf{M}$ within the patterned domain. (g, h) Gradual control of complementarity of the magnetic domains within the "CU"-shaped region and in its surrounding. The cell thickness is 30μm and applied fields are ≈3 mT. The red-light illumination in POM imaging prevents unintended isomerization of LC molecules.

In addition to patterning of $\mathbf{M}$ in cells with uniform director $\mathbf{n_0}$, co-patterning of spatially non-uniform magnetization $\mathbf{M(r)}$ and director $\mathbf{n(r)}$ fields can be achieved too. To demonstrate this, we use optical control of surface boundary conditions for $\mathbf{n(r)}$ and the non-isomerizing 5CB-based FLCC. We design a defect-containing pattern of $\mathbf{n(r)}$ that contains two half-integer disclinations in $\mathbf{n(r)}$, both orthogonal to cell-substrates (Fig. 3a) [8]. These disclinations in the nematic bulk are line defects with singular cores, around both of which $\mathbf{n(r)}$ rotates by $\pi$ in opposite directions when one circumnavigates the core of the defect once, defining the respective disclination winding numbers of ±1/2 [8]. Although these are the only patterned defects in $\mathbf{n(r)}$, due to its polar nature, the magnetization field directed so that $\mathbf{M(r)}\|\mathbf{n(r)}$ is topologically required to contain a wall defect connecting them because standalone half-integer defect lines are not allowed in vector fields like $\mathbf{M(r)}$ (Fig. 3b) [8]. To observe what happens when defects patterned in $\mathbf{n(r)}$ are incompatible with the polar nature of $\mathbf{M(r)}$, we realize the field configurations shown in Fig. 3. The light-tunable alignment of $\mathbf{n}(\mathbf{r})$ in the FLCC bulk is implemented using confining glass plates with the inner surfaces spin-coated by azobenzene-containing alignment layers of PAAD-22 (Beam Co.). Before filling the cells with the FLCC, far-field uniform planar alignment was first defined by exposing the cell to linearly polarized broadband white light for ~5 min. Then, the spatially non-uniform boundary conditions for $\mathbf{n}(\mathbf{r})$ on the substrates within a cell area of interest were patterned through a sequential illumination process using light of varying linear polarization, as illustrated in Fig. 3a. Taking into account that the azobenzene molecular units of PAAD-22 orient orthogonally to the linear polarization of the illumination light, we defined discrete regions of the cell with orientations of the easy axis for $\mathbf{n}(\mathbf{r})$ as shown in Fig. 3a. A large number of discrete easy axis orientation regions, along with the elastic cost of producing sharp distortions on small lengthscales, assured relatively continuous $\mathbf{n}(\mathbf{r})$ of the intended disclination pair embedded in the uniform far-field $\mathbf{n}_0$ (Fig. 4). A salient feature of cell regions containing this director pattern is that a wall defect connecting the two vertical defect lines is apparent even at no applied fields (Fig. 4a,d,g). At the wall defect, the orientation of $\mathbf{M}(\mathbf{r})$ flips to an opposite in-plane orientation, as depicted in Fig. 3b, although $\mathbf{n}(\mathbf{r})$ stays continuous (Fig. 3a) because half-integer defects are allowed in the $\mathbf{n}(\mathbf{r})$ line field but not in the $\mathbf{M}(\mathbf{r})$ vector field.

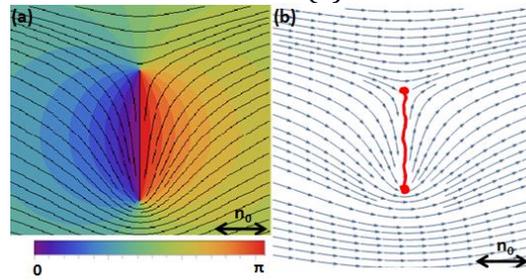

Fig. 3. Principles of optical patterning and structure of singular defects in $\mathbf{n}(\mathbf{r})$ and $\mathbf{M}(\mathbf{r})$. (a) The two-dimensional $\mathbf{n}(\mathbf{r})$ of a half-integer disclination pair in a uniform far-field background $\mathbf{n}_0$. The two singularities in $\mathbf{n}(\mathbf{r})$ correspond to -1/2 (top) and +1/2 (bottom) $\mathbf{n}(\mathbf{r})$ defect lines orthogonal to the plane of the figure. To photo-pattern boundary conditions for $\mathbf{n}(\mathbf{r})$ on the cell substrates treated by PAAD-22 and define this defect pair in $\mathbf{n}(\mathbf{r})$ throughout the cell, we used discrete orientations of linear polarizations of illumination light, as depicted using the color-angle scheme, where orientations of $\mathbf{n}(\mathbf{r})$ relative to $\mathbf{n}_0$ are given in radians with respect to the vertical edge of the figure. (b) The corresponding $\mathbf{M}(\mathbf{r})$ contains a wall defect connecting the two disclinations in $\mathbf{n}(\mathbf{r})$, which is caused by the polar nature of $\mathbf{M}$. The singular wall in $\mathbf{M}(\mathbf{r})$ and line defects in $\mathbf{n}(\mathbf{r})$ are translationally invariant in the direction of sample thickness normal to the figure and are depicted using red wiggly line and filled circles, respectively.

Although, in principle, simultaneous patterning of both $\mathbf{M}(\mathbf{r})$ and $\mathbf{n}(\mathbf{r})$ can be achieved by combining the two approaches described above, it is interesting to note the evolution of domain structures in $\mathbf{M}(\mathbf{r})$ when only the director is patterned (Fig. 4). The $\mathbf{M}(\mathbf{r})$ within magnetic domains in the patterned region follows the spatially varying $\mathbf{n}(\mathbf{r})$ within the magnetic domains and behaves discontinuously (flipping to opposite domains) at the inter-domain walls. At applied magnetic fields, the polydomain nature of the FLCC interplays with the topologically required wall connecting the half-integer defect lines in **n(r)**, causing a complex pattern of domains and wall defects inter-spacing them, which slowly evolve with time and strongly depend on both the direction and strength of **B**. Interestingly, the width of wall defects within the regions of distorted **n(r)** is often larger than that in regions of uniform director (Fig. 4). To uncover the nature of magnetic inter-domain walls in FLCCs, we used dark field microscopy observations that reveal both locations and orientations of individual nanoplatelets (supplementary S1) at no fields and when **B** at different orientations selectively switches the domains of opposite **M**. Unlike in conventional magnetic systems, where magnetic domains are typically separated by the so-called Bloch or Néel walls [8] with continuous albeit localized solitonic deformations of **M(r)**, magnetization at the inter-domain walls of FLCCs is not defined, so that they are singular in nature. This is because **M(r)** and **n(r)** are strongly coupled, so that the solitonic deformations of **M(r)** between domains would be costly in terms of the corresponding elastic deformations of **n(r)**. Instead, the domain walls in the FLCC have uniform director but undefined **M(r)**, so that there is no associated elastic free energy cost due to such walls. At applied fields, the inter-domain walls can be partially deprived of nanoparticles (supplementary video S1) and ranging in width from the average spacing between individual nanoparticles to ~1μm, as determined by colloidal interactions between nanoplatelets with differently oriented dipole moments of the neighboring domains of opposite **M**‖**n**₀. When **n(r)** ‖**M(r)** within the domains is distorted, this interplay is further altered by the energetic cost of elastic distortions (Figs. 4). Although the domain wall defects in FLCCs are singular in **M(r)**, the one-, two- and three-dimensional twisted solitons were also observed in the chiral counterparts of FLCCs, in which twisted **n(r)** ‖**M(r)** structures are promoted by the chiral nature of the LC host fluid [16].

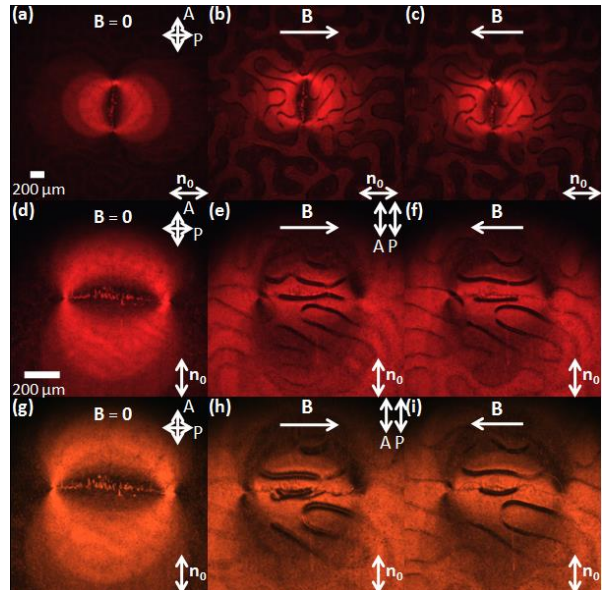

Fig. 4. POM images of FLCCs with photo-patterned defects. (a) A micrograph of the structure corresponding to **n(r)** shown in Fig. 3, with the wall in **M(r)** visible as a bright jiggling line. An in-plane **B** along opposite directions marked in (b, c) switches complementary domains with anti-parallel non-uniform **M(r)**. (d-i) POM images of the FLCC cell with a disclination pair. The inter-disclination and other **M**-wall

defects separating magnetic domains are clearly visible. Polarizing micrographs (a, b, c, d, g) were taken with crossed polarizer "**P**" and analyzer "**A**" while the rest were obtained with parallel **P** and **A**, as shown using white double arrows.

Unlike conventional LCs, which exhibit a response to magnetic fields that is much less facile than the response to electric fields, FLCCs are switched at $\approx 1$ mT fields and exhibit polar response. Although either only magnetically uniform or only random polydomain FLCC samples were studied previously [4,7,16], our work demonstrates that the magnetic domains can be patterned on mesoscopic length-scales in many different ways, thus allowing for engineering the macroscopic response of such materials to magnetic fields. Since this patterning is achieved, in a re-configurable fashion, using low-intensities of unstructured light, our findings may enable designs and implementations of mesostructured soft matter composites with unusual properties and facile pre-defined responses to weak external stimuli. Beyond the potential practical uses, the experimental framework we have developed may allow one to study how the polar nature of ordering in soft matter gives rise to new types of topological defects and solitons. In addition to small-molecule liquid crystalline hosts of magnetic nanoparticles, FLCC ordering can potentially be patterned in polymeric, elastomeric, and other soft matter systems.

To conclude, we have demonstrated optical patterning of magnetic domains and defects in FLCCs. Because of the facile magnetic switching with easily configurable domain polarity and elastic distortion patterning, our experimental framework may lead to realizing structures and composites well beyond the ones presented in this letter. The introduction of chirality and particle-like excitations can enable magnetic control of complex optical phenomena, such as nematicon formation and deflection [17], selective reflection tuning [18,19], non-contact manipulation of defects [20,21], etc. Moreover, polymer or elastomer networks in FLCCs may yield novel mechanical behavior of the ensuing soft solids, such as programmable strain biases and selective mechanical actuation [12,14].

We acknowledge discussions with N. Clark, W. Gannett, M. Keller, P. Ackerman, A. Martinez, T. White, and Y. Zhang. This research was supported by the NSF grant DMR-1420736.

## Movie Caption

**Movie S1.** A dark-field microscope movie shows scattering of ~0.1 wt.% magnetic nanoplates within FLCC domains in a 15.5 μm-thick LC cell with homeotropic surface anchoring and $\mathbf{n_0}$ obtained using a $100\times$ objective. At first, no field was applied on the cell, then a magnetic field of ~2 mT was applied along $\mathbf{n_0}$, following by switching the magnetic field by $180°$. Finally the magnetic field was removed and the FLCC relaxed to the original state.